\documentclass[11pt,a4paper]{article}

\usepackage[utf8]{inputenc}

\usepackage{cite}

\usepackage{hyperref}
\usepackage{lipsum}
\usepackage{amsmath, amssymb,slashed,mathrsfs, bbm, braket, color}

\numberwithin{equation}{section}

\def\nn{\nonumber}

\def\im{{\rm i}}

\def\L{\mathcal L}

\def\a{\alpha}
\def\b{\beta}
\def\g{\gamma}

\def\x{\varphi}
\def\y{ \chi}
\def\z{ \zeta}

\newcommand{\be}[1]{\begin{equation}#1\end{equation}}
\newcommand{\ba}[1]{\begin{align}#1\end{align}}
\newcommand{\p}[1]{\left(#1\right)}
\newcommand{\wt}[1]{\widetilde{#1}}
\newcommand{\comm}[1]{\left[#1\right]}
\newcommand{\acomm}[1]{\left\{#1\right\}}
\newcommand{\f}[2]{\frac{#1}{#2}}
\newcommand{ \bb}[1]{{\mathbb #1}}

\begin{document}
\begin{titlepage}
\begin{flushright} 
UUITP -- 63/18
\end{flushright}
\vskip 1.5in
\begin{center}
{\bf\Large{    Supersymmetric Localization of Refined Chiral Multiplets on Topologically Twisted $ H^2 \times S^1$ }}
\vskip
0.5in { {Antonio Pittelli }
} \vskip 0.5in {\small{ 
\emph{Department of Physics and Astronomy, Uppsala University,\\ Box 516, SE-75120 Uppsala, Sweden} 
}
}
\end{center}
\vskip 0.5in
\baselineskip 16pt

\begin{abstract}

We derive the partition function of an $\mathcal N=2$ chiral multiplet on topologically twisted $H^2\times S^1$. The chiral multiplet is  coupled to a background vector multiplet encoding a real mass deformation. We consider an $ H^2\times S^1$   metric containing  two parameters: one is  the $S^1$ radius,  while   the other gives   a fugacity $q$ for the angular momentum on $H^2$. The computation is carried out by means of  supersymmetric  localization, which provides a  finite answer  written in terms of  $q$-Pochammer symbols and   multiple Zeta functions. Especially,  the partition function of normalizable fields   reproduces three-dimensional holomorphic blocks.  

\end{abstract}

\date{}
\end{titlepage}

\tableofcontents

\newpage 

%%%%%%%%%%  Core     %%%%%%%%%%%

 \section{Introduction and Conclusions}

  Localization  techniques have considerably improved our understanding of   quantum field theory as they allow for the exact  computation of interesting  physical observables. They were  first applied to topological field theories  \cite{Witten:1992xu} and then extended to supersymmetric gauge  theories in diverse dimensions  \cite{Nekrasov:2003af, Nekrasov:2003rj, Pestun:2007rz, Kapustin:2009kz, Marino:2009jd}. A consistent definition of supersymmetric quantum field theories  on curved manifolds   \cite{Festuccia:2011ws, Dumitrescu:2012ha, Closset:2012ru, Closset:2013vra, Closset:2014uda} was  crucial in enlarging  the applicability of  localization, which keeps producing  several    non-perturbative results such as  tests of the AdS/CFT correspondence and other supersymmetric dualities   \cite{Marino:2009jd, Benini:2015eyy, Benini:2016rke, Cabo-Bizet:2017jsl, Hosseini:2017mds, Hosseini:2017fjo , Benini:2017oxt,  Cabo-Bizet:2018ehj, Choi:2018hmj, Benini:2018ywd }.  The literature on the subject is gargantuan and we refer the reader to the recent review \cite{Pestun:2016zxk} and to the references therein. 
  
  Localization on compact manifolds is largely investigated, see e.g. \cite{Benini:2012ui, Benini:2015noa, Benini:2016hjo, Alday:2013lba, Closset:2013sxa, Assel:2014paa, Closset:2015rna, Festuccia:2016gul, Polydorou:2017jha, Festuccia:2018rew}. Less understood is    localization on compact manifolds with boundary \cite{Hori:2013ika,Yoshida:2014ssa, Gava:2016oep, Bawane:2017gjf}; even less the case of non-compact hyperbolic  manifolds \cite{Aharony:2015hix, Bonetti:2016nma, David:2016onq, Assel:2016pgi, Cabo-Bizet:2017jsl, David:2018pex}.  In this paper we  localize the partition function $Z_{\rm chi}$ of    a chiral multiplet with arbitrary R-charge $r$ on   $H^2\times S^1$. The model is topologically twisted as the R-symmetry background is chosen in order to cancel  the spin connection, allowing  for covariantly constant Killing spinors.   We consider a chiral multiplet  coupled to a background vector multiplet inducing a real  mass deformation. In analogy with \cite{Benini:2015noa}, once specified the action functional $S_{\rm chi}\comm{\Psi}$ and the boundary conditions $\Psi_\partial$ for the chiral multiplet fields $\Psi$,   the observable $Z_{\rm chi}$ admits a path-integral representation as well as  a canonical quantization definition in terms of a trace over the Hilbert space ${\mathcal H\comm{H^2}}$ of states on $H^2$:
  \ba{\label{eq: basicdefs}
  Z_{\rm chi} = \int_{\Psi_\partial}\comm{d\Psi} e^{ - S_{\rm chi}\comm{\Psi}}={\rm Tr}_{\mathcal H\comm{H^2}}\p{-1}^{\mathscr F} e^{2\pi \im\,  \mathscr H} ={\rm Tr}_{\mathcal H\comm{H^2}}\p{-1}^{\mathscr F} q^{  P_\chi} \,  t^{  J_F}  ~ , 
  }
  where $\mathscr F$ is the fermion number and   $\mathscr H$  the translation operator along the $S^1$ orthogonal to $H^2$. The second equality descends from the supersymmetry algebra $Q^2 = - \mathscr H + \alpha \, P_\chi - \im \, u \, J_F $, with $P_\chi$ being the (R-symmetry twisted) angular momentum on $H^2$, $J_F$ generating flavor  symmetries  and $q=e^{2\pi \im \alpha}$, $ t=e^{2\pi u}$  being   fugacities thereof.  The case $\alpha=0$, $q=1$ was studied  in  \cite{Cabo-Bizet:2017jsl}.
  
  As in \cite{Assel:2016pgi}, the answer for $Z_{\rm chi}$ strongly depends on boundary conditions and on the normalizability of states contributing to the partition function.  Indeed, if we include both normalizable and non-normalizable contributions, we obtain
   \ba{\label{eq: nonnormresult}
&  r\neq1 :    Z_{{\rm chi}}=e^{\im\pi\, \mathcal A_{\rm chi}}\f{(t \,q^{1-\f r2};q)}{(t^{-1}\, q^{\f r2};q)}\,, \qquad  r=1 :  Z_{\rm chi} = 1\,; & (z;x):=\prod_{m\geq0}(1-z\,x^m) \,,
  }  
  with  $u = L\,\b\p{\sigma+\im\,  v'}$. Here, $\sigma$ is a real mass deformation and  $v'$ a  particular component of  a background field corresponding to a flavour symmetry ${\rm U}(1)_F$. Moreover, $L$ is the $H^2$ radius, $\b$ the ratio between the $S^1$ radius and $L$   and $\a\in \bb R$   a real parameter deforming the $H^2\times S^1$ metric.  The phase factor $\mathcal A_{\rm chi}$ is given in terms of double zeta functions, 
     \be{
     \mathcal A_{\rm chi}=\mathcal A_{ B}- \mathcal A_{\phi} \,,\qquad  \mathcal A_{B}=\zeta_2(0,\a\,- \f{\a r}2+\im u|1,\a) \,, \qquad \mathcal A_{\phi}=\zeta_2(0,\f{\a r}2-\im u|1,\a)\,.
     }
     In particular, $Z_{{\rm chi}}=1$ at $r=1$ because no state satisfies  supersymmetric boundary conditions for that specific value of the R-charge. Forby, the absolute value of $Z_{{\rm chi}}$ for $r\neq1$ is the plethystic exponential \cite{Feng:2007ur} of a  \emph{single letter partition function} $f_r(t,q)$:
       \be{
    f_r(t,q)=\f{t^{-1}\,q^{\f r2}-t\,q^{1-\f r2}}{1-q}\,,\qquad Z_{\rm chi}=e^{\im \,\mathcal A_{\rm chi}}\,{\rm P.E.}[f_r(t,q)]\,.
    }
    If we  shrink the $S^1$ radius  by taking the limit $\b\to0$, the single letter reduces to   $f_r(1,q)=({q^{\f r2}-q^{1-\f r2}})/\p{1-q}$.  Notice that $Z_{{\rm chi}}$ does not depend  on the $S^1$ radius  $\b$   in absence  of   background vector multiplets,   becoming  a continuous function of $r$. 
    
    On the other hand, if we  exclude non-normalizable contributions,   ${  Z_{\rm chi}}$ reads  
    \be{\label{eq: secondresult}
  r<1 :   Z_{\rm chi}=e^{\im\pi\, \mathcal A_B}{(t \,q^{1-\f r2};q)} \,,\qquad  r=1 :  Z_{\rm chi} = 1\,,\qquad r>1 : Z_\phi=\f{e^{\im\pi\, \mathcal A_\phi} }{(t^{-1}\, q^{\f r2};q)} \,.
  }
    Equations (\ref{eq: secondresult}) are reminiscent of what happens in topologically twisted theories on $\mathcal M^2\times S^1$, where the R-symmetry background produces Landau levels for a quantum mechanics on $S^1$ \cite{Almuhairi:2011ws, Kutasov:2013ffl, Benini:2015noa, Cabo-Bizet:2017jsl}. 
    
    For $r\neq1$, the partition functions  $Z_{\rm chi}$ in (\ref{eq: secondresult}) reproduces   three-dimensional holomorphic blocks \cite{Beem:2012mb, Nieri:2015yia}, also obtained  by performing supersymmetric  localization on  $D^2\times S^1$ \cite{Yoshida:2014ssa}.  In light of (\ref{eq: secondresult}), including non-normalizable contributions in the partition function calculation  amounts to trivially gluing together $Z_{\rm chi}(r<1)$ and $Z_{\rm chi}(r>1)$. This procedure  yields  the partition function on $S^2 \times S^1$,  explaining the accidental coincidence between   the  3d superconformal index of a chiral multiplet with arbitrary R-charge \cite{Imamura:2011su} and the (a priori different) topologically twisted index (\ref{eq: nonnormresult}).

    As we have already mentioned, the value $r=1$ is special   from the viewpoint of boundary conditions as well. Indeed, an R-charge $r>1$ implies  Dirichlet boundary conditions on the  fields contributing to $Z_{{\rm chi}}$; namely,  the scalars $\phi,\wt \phi$ are supposed to vanish at the (conformal) boundary. On the other hand, $r<1$ requires Robin boundary conditions, meaning that derivatives of $\phi,\wt \phi$   go to zero  at the boundary. The case $r=1$ does not correspond to any set of BPS boundary conditions and, in fact, there are no fields contributing to $Z_{{\rm chi}}$ non-trivially for $r=1$.  
 
  %%%%%%%%%%%%%%%%%%%%%%
 
 \subsection{Outlook}
 
 In this paper we studied an $\mathcal N=2$ chiral multiplet on topologically twisted $ H^2\times S^1$ coupled to a background vector multiplet incorporating a real mass deformation. It would be very interesting to generalize  the results of the present work by including dynamical vector multiplets, Chern-Simons terms as well as general  BPS observables. This would provide  a complete study of gauge theories on $H^2\times S^1$, helping out to clarify universal  features  of supersymmetric theories on non-compact manifolds, also unveiling  possible dualities intertwining them.
 
Furthermore, it would be intriguing  to apply a similar analysis to gauge   theories defined on    higher dimensional non-compact manifolds. This not only would be fascinating  per se, but   should also shed a new light on our findings concerning  matter  multiplets on $H^2\times S^1$. 

 Finally,   it would be compelling   to explore  the link  between partition functions on $H^2\times S^1$, the half-index on $D^2\times S^1$ and 3d holomorphic blocks. In particular, a rigorous  interpretation of (\ref{eq: secondresult}) in terms of a quantum mechanics for states on   $H^2$   would be desirable\footnote{We thank Pietro Longhi for raising this point.}.

  %%%%%%%%%%%%%%%%%%%%%%
  
   \subsection{Outline}
   
   In Section 2 we describe the geometry of topologically twisted $H^2\times S^1$, constructing the corresponding Killing spinors. In Section 3 we write down the supersymmetry transformations and action for an $\mathcal N=2$ chiral multiplet  coupled to a background vector multiplet. We shall also introduce twisted fields, which simplify the localization computation, and discuss the asymptotic boundary conditions. Eventually, Section 4 contains the computation of the one-loop determinant for the chiral multiplet on twisted $H^2\times S^1$.
 
  %%%%%%%%%%%%%%%%%%%%%%
 
 \subsection{Acknowledgements}
 
 I am very grateful to Guido Festuccia, Pietro Longhi, Fabrizio Nieri and  Achilleas Passias for precious  discussions and for comments on the draft. The work of AP is supported by the ERC STG Grant 639220.
 
 %%%%%%%%%%%%%%%%%%%%%%
  %%%%%%%%%%%%%%%%%%%%%%
  
   \section{Background Geometry}
   
    \subsection{Metric and Killing Spinors}
     
    We use the conventions of \cite{Assel:2016pgi}, which are the same as the conventions of \cite{Closset:2012ru} apart from a sign in the definition of the spin connection. Let us consider $H^2\times S^1$ with line element
    \be{
    ds^2=L^2\comm{d\eta^2 + \sinh^2\eta\p{d\y+\a\, d\x}^2}+L^2\b^2 d\x^2, \qquad \eta\in \bb R^+\,,\quad \y,  \x\in\comm{0,2\pi} \,,
    }
    where $\alpha,\beta\in\bb R$. Consequently, the orthonormal frame $e^a $ is
    \be{
    {e^1}=L \, d\eta,\qquad     {e^2}_\y=L \sinh\eta \, d\y+L\, \a \sinh\eta\,d\x ,\qquad  {e^3}=L \,\b\,d\x ,   
    }
    In our conventions, the Ricci scalar of $H^2\times S^1$ is $R=-2/L^2$. The Killing spinor equations for a three-dimensional manifold with $\mathcal N=2$ supersymmetry read 
 \ba{\label{eq: kse}
 &\nabla_\mu\z-\im A_\mu \z = -\f H2\gamma_\mu\z - \im V_\mu\z -\f12\epsilon_{\mu\nu\rho}V^\nu\gamma^\rho\z\,,\nn\\
 &\nabla_\mu\wt\z+\im A_\mu \wt\z = -\f H2\gamma_\mu\wt\z + \im V_\mu\wt\z +\f12\epsilon_{\mu\nu\rho}V^\nu\gamma^\rho\wt\z\,.
 }
 If we choose the background fields
 \be{
 A=\f12\cosh\eta\p{d\y + \alpha\,d\x}\,,\qquad  H=0\,,\qquad V=0\,,
 }
we find that the  spinors
 \ba{\label{eq: h2s1ks}
 \z_\a= \f1{\sqrt2} \,(1 ,\im)_\a\,, \qquad  \wt  \z_\a=- \f1{\sqrt2}\,( \im , 1 )_\a\,,  
 }
 solve (\ref{eq: kse}). The Killing spinors $\z,\wt \z$ in  (\ref{eq: h2s1ks}) have R-charges $1, -1$ respectively. They satisfy $\z \wt \z=-\wt \z \z =+1$ as well as $\zeta^\dagger=-\wt \z$, implying   $|\z|^2=\z^\dagger\z= |\wt \z|^2=\wt \z^\dagger \wt \z=+1$. 
 
 %%%%%%%%%%%%%%%%%%%%%%
     
        \subsection{Three-Dimensional Frame}

 The Killing spinors  $\z,\wt \z$ allow for constructing  the bilinears  
 \be{\label{eq: kpptilde}
 K^\mu=\z\g^\mu\wt \z\,,\qquad  P^\mu=\z\g^\mu  \z\,,\qquad  \wt P^\mu=\wt \z\g^\mu\wt \z\,.
 }
The vectors $ K^\mu,P^\mu,\wt P^\mu $ have R-charges $0,2,-2$   and fulfil 
 \be{
 g^{\mu\nu}= K^\mu K^\nu-P^{(\mu}\wt P^{\nu)}\,,\quad K_\mu K^\mu = 1\,,\quad \wt P_\mu P^\mu = -2\,,\quad \p{K_\mu}^*=K_\mu\,,\quad \p{P_\mu}^*=-\wt P_\mu\,.
 }
By contracting  (\ref{eq: kpptilde}) with $\partial_\mu$, we obtain a representation in terms of Lie derivatives $  \L_K = K^\mu\, \partial_\mu $, $   \L_P =     P^\mu\,\partial_\mu $ and $  \L_{\wt P} =      \wt P^\mu\,\partial_\mu$:
     \be{
    \L_K   =- \f1{L\,\b} \p{\a\,\partial_\y-\partial_\x} \,,\qquad  \L_P  =\f1L\p{\im\,\partial_\eta+\f1{\sinh\eta} \partial_\y}  \,,\qquad  \L_{\wt P}    = \f1L\p{\im\,\partial_\eta-\f1{\sinh\eta} \partial_\y}  \,.  
    }
    %Alternatively, contracting (\ref{eq: kpptilde}) with $dx^\mu$ provides a  representation in terms of one-forms $\gen K= K_\mu \,dx^\mu$ , $ \gen P =     P^\mu \,dx^\mu$   and  $ \wt{\gen P} =      \wt P_\mu\, dx^\mu $:
    %\be{
    % \gen K = L\,\b\,d\x  \,,\qquad   \gen P   = L\comm{\im\,d\eta+\sinh\eta\p{d\y+\a\,d\x}}    \,, \qquad  \wt{\gen P}   =  L\comm{\im\,d\eta-\sinh\eta\p{d\y+\a\,d\x}}   \,.  
   % }
   Especially,  the parameter $\alpha$ deforms $\mathcal L_K$ by a term proportional to $\partial_\chi$, where the latter is the angular momentum operator on $H^2$.

 %%%%%%%%%%%%%%%%%%%%%%
  %%%%%%%%%%%%%%%%%%%%%%
  
   \section{Chiral Multiplet   on $H^2\times S^1$}
   
   \subsection{Supersymmetry Transformations and Action}
   
  The supersymmetric  transformations for a chiral multiplet  $\p{\phi,\psi,F}$ of R-charge $r$ on $H^2\times  S^1$ with respect to the supercharge  $\delta = \delta_\z + \delta_{\wt \z}$  are \cite{Closset:2012ru}
 \ba{
 & \delta\phi = \sqrt2\,\z\psi\,,\nn\\
 & \delta\psi=\sqrt2\, \z F +\im\sqrt2 \,\sigma\,\phi\wt \z - \im\sqrt2\,\gamma^\mu\wt \z\mathcal D_\mu\phi\,,\nn\\
 & \delta F =-\im\sqrt2\,\sigma\,\wt\z\psi - \im\sqrt2\,\mathcal D_\mu\p{\wt\z \gamma^\mu\psi}\,,
 }
 where we introduced the covariant derivative $\mathcal D_\mu=\nabla_\mu-\im\, q_R\p{A_\mu-\f12 V_\mu}-\im\, v_\mu$. Here, $q_R$ is the R-charge, $\sigma$   a constant scalar   encoding a real mass deformation and  $v_\mu$ a   background gauge field   corresponding to a flavor symmetry ${\rm U}(1)_F$. Similarly,   we can write down the supersymmetry transformations for an anti-chiral multiplet  $\p{\wt \phi,\wt \psi,\wt F}$ of R-charge $-r$:
  \ba{
 & \delta \wt \phi = - \sqrt2\,\wt \z \wt \psi\,,\nn\\
 & \delta \wt \psi=\sqrt2\, \wt \z \wt  F -\im\sqrt2 \,\sigma\,\wt \phi  \z + \im\sqrt2\,\gamma^\mu \z\mathcal D_\mu\wt\phi\,,\nn\\
 & \delta \wt F =-\im\sqrt2\, \sigma\, \z\wt \psi - \im\sqrt2\,\mathcal D_\mu\p{ \z \gamma^\mu\wt\psi}\,.
 }
The supersymmetric variations $\delta_\z,\delta_{\wt \z}$ are nilpotent, while   $\delta$ squares to an isometry of the background $\mathcal L_K$ plus a central charge given by the background fields:  
\be{
\delta^2=\acomm{\delta_\z,\delta_{\wt \z}}=-2\im\,\mathcal L_K+2\im\p{\sigma+\im\,K^\mu\, v_\mu}\,.
} 
The action for the above chiral multiplet is given by integrating over $ H^2\times S^1$ the following Lagrangian:
\ba{\label{eq: chirallagh2s1}
\mathcal L_{\rm chi}=\mathcal D^\mu\wt \phi\,\mathcal D_\mu\phi+\p{\sigma^2-\f r{2 L^2}}\wt\phi\phi-\wt FF-\im\,\wt \psi\gamma^\mu\mathcal D_\mu\psi-\im\,\sigma\,\wt\psi\psi\,.
}

 %%%%%%%%%%%%%%%%%%%%%%
 
    \subsection{Twisted Fields}
    
    Let us introduce the twisted fields $B,C,\wt B,\wt C$, which are  Gra\ss mann-odd scalars of R-charge $\p{r-2,r,2-r,-r}$ defined as \cite{Assel:2016pgi}
    \ba{
    & B =\wt\z\psi\,,\qquad C=\z\psi\,,\qquad \wt B = \z \wt\psi\,,\qquad C=-\wt\z\wt\psi\,, \nn\\
    & \psi=\z B + \wt\z C\,,\qquad  \wt\psi=\wt\z \wt B +  \z \wt C\,.
    }
    The non-trivial supersymmetric variations of $\p{\phi,B,C,F}$ read
    \ba{
    & \delta_\z\phi=\sqrt2 C\,,\qquad \delta_{\wt \z} C = -\im\sqrt2\hat\L_K\phi+\im\sqrt2\sigma\phi \,,\nn\\
    & \delta_\z B=\sqrt2 F\,,\qquad \delta_{\wt \z}B=\im\sqrt2\hat\L_{\wt P}\phi\,,\nn\\
    & \delta_{\wt \z} F=-\im\sqrt2\hat\L_KB+\im\sqrt2\sigma B-\im\sqrt2\hat\L_{\wt P}C\,,
    }
    while those of $\p{\wt \phi,\wt B,\wt C,\wt F}$ are
    \ba{
    & \delta_{\wt\z}\wt\phi=\sqrt2 \wt C\,,\qquad \delta_{  \z} \wt C = -\im\sqrt2\hat\L_K\wt \phi-\im\sqrt2\sigma \wt \phi \,,\nn\\
    & \delta_{\wt\z} \wt B=\sqrt2 \wt F\,,\qquad \delta_{  \z}\wt B=\im\sqrt2\hat\L_{  P}\wt \phi\,,\nn\\
    & \delta_{  \z}\wt  F=-\im\sqrt2\hat\L_K\wt B-\im\sqrt2\sigma\wt  B-\im\sqrt2\hat\L_{  P}\wt C\,.
    }
    Here,   \emph{hatted} Lie derivatives are covariant, for example $\hat{\mathcal L}_X=X^\mu\mathcal D_\mu$. Via  twisted fields we can  write down the     deformation term
    \be{\label{eq: defterm}
    \mathcal V_{\rm chi}=\f12\comm{\p{\delta_\z B}^\ddagger B+\p{\delta_\z \wt B}^\ddagger \wt B+ \p{\delta_\z \wt C}^\ddagger \wt C}=\f1{\sqrt2}\comm{\im\,\wt B \hat\L_{\wt P}\phi+\im\,\wt C\p{\hat \L_K\phi+\sigma\phi}-\wt F B}\,,
    }
    where we used the reality conditions $\phi^\ddagger=\wt \phi $ and $ F^\ddagger=-\wt F$, the involution $\ddagger$ acting as complex conjugation upon $c$-numbers. The variation of $\mathcal V_{\rm chi}$ with respect to the supercharge $\delta_\z$ yields the Lagrangian 
    \ba{\label{eq: twistedchirallagh2s1}
    \L_{\rm chi}' &=\L_K\wt \phi\L_K\phi-\L_P\wt \phi\L_{\wt P}\phi+\sigma^2\wt \phi\phi-\wt FF+\nn\\
    & -\im\,B\L_K\wt B -\im\,\wt C\L_K C -\im\,B\L_P\wt C -\im\,\wt B\L_{\wt P} C +\im\,\sigma\p{\wt B B+C \wt C}\,,
    }
    coinciding with (\ref{eq: chirallagh2s1}) up to total derivatives. By construction, (\ref{eq: twistedchirallagh2s1}) is supersymmetric under both $\delta_\z$ and $\delta_{\wt \z}$ without imposing any boundary condition\footnote{Indeed, $\delta_\z \L_{\rm chi}' =0$, while  $\delta_{\wt \z} \L_{\rm chi}' =-2\im\,\L_K\mathcal V_{\rm chi}\to \delta_{\wt \z} S_{\rm chi}'=0  $ as $\mathcal L_K$ is parallel to the boundary.}.   
    
     %%%%%%%%%%%%%%%%%%%%%%
     
        \subsection{Boundary Conditions}
        
        If we use the Lagrangian (\ref{eq: twistedchirallagh2s1}), we find that the equations of motion of $B,F,\wt B,\wt F$   generate bulk terms only. Instead, the equations of motion of $\phi, C ,\wt \phi,\wt C$ give\footnote{An analogous  approach was   used in the study of supersymmetric theories on Euclidean $H^3$ \cite{Assel:2016pgi} and to derive dual boundary conditions in three-dimensional superconformal field theories \cite{Dimofte:2017tpi}.}
        \ba{\label{eq: twistedchiraleomh2s1}
    \delta_{\rm eom}S_{\rm chi}' &=\delta_{\rm eom}\int_M d^3x\,\sqrt g\, \L_{\rm chi}'\nn\\
    & =\p{\rm bulk} -\int_{  M} d^3x\,\sqrt g\,\comm{ \L_P( \delta \wt \phi\L_{\wt P}\phi)  +\L_{\wt P}(\delta\phi  \L_P\wt \phi ) +\im\,\L_P(B\delta \wt C) +\im\,\L_{\wt P}(\wt B\delta  C) }\,,
    }
    where $M=H^2\times S^1$. The $\p{\rm bulk}$ terms vanish by the equations of motion, while the boundary terms disappear if we impose at the conformal boundary either   Dirichlet boundary conditions, $\phi=\wt \phi= C =\wt C = 0$, or  Robin boundary conditions $\mathcal L_{\wt P}\phi=\mathcal L_{  P}\wt \phi= B =\wt B = 0$. If we choose to leave the field variations $\delta\phi,\delta C, \delta\wt \phi,\delta \wt C$ free to oscillate at the conformal boundary, the action $S'_{\rm chi}$ forces us to impose Robin. As we shall see in the next section, asymptotic boundary conditions will constrain the R-symmetry of the modes contributing to the one-loop determinant of the partition function.

 %%%%%%%%%%%%%%%%%%%%%%
  %%%%%%%%%%%%%%%%%%%%%%
  
   \section{Localization}

   \subsection{BPS Locus}
   
   The deformation term (\ref{eq: defterm}) leads to the Lagrangian  (\ref{eq: twistedchirallagh2s1}), whose bosonic part is positive definite. The saddle point configurations of the path integral are then obtained by solving  the BPS equations
   \be{
    \delta_\zeta B=\delta_\zeta \wt B= \delta_\zeta C=\delta_\zeta \wt C=0\,.
   }
   These constraints immediately imply $F=\wt F=0$. Furthermore,   periodicity along $(\x,\y)$ directions yield $\phi=\wt \phi=0$. We then find the  trivial locus  $\phi=\wt \phi=F=\wt F=0$.
   
 %%%%%%%%%%%%%%%%%%%%%%
 
    \subsection{One Loop Determinant}
    
    We compute the one loop determinant by means of the unpaired eigenvalues method, see {e.g.} \cite{Hama:2011ea, Closset:2013sxa, Assel:2016pgi, David:2018pex}.  This exploits two main facts: first,  $\hat\L_P, \hat\L_{\wt P}$ commute with the operator   $\delta^2$, whose functional  determinant  provides the chiral multiplet  partition function $Z_{\rm chi}$. Second, $\hat\L_P, \hat\L_{\wt P}$ map to each other bosonic and fermionic modes. As a result, the neat contribution to $Z_{\rm chi}$ is  given by modes belonging to the kernels of $\hat\L_P, \hat\L_{\wt P}$:
    \be{\label{eq: oneloopdetkernelcokernel}
    Z_{\rm chi}=\f{\det_{{\rm Ker}\,\hat\L_P} { \delta^2}}{ \det_{{\rm Ker}\,\hat\L_{\wt P}}{  \delta^2}  } \;.
    }
 In our setup, such modes  have the form
 \ba{
 & {\rm Ker}\,\hat\L_{\wt P} \,:  \phi_{m_\x,m_\y}=e^{\im \,m_\x\, \x+\im \,m_\y\, \y} \p{\tanh\f\eta2}^{m_\y}\p{\sinh\eta}^{-\f{r}2}\,,\nn\\
 & {\rm Ker}\,\hat\L_{  P}\,:  B_{n_\x,n_\y}=e^{\im \,n_\x\, \x+\im \,n_\y\, \y} \p{\coth\f\eta2}^{n_\y}\p{\sinh\eta}^{\f{r-2}2}\,,
 }
 with $m_\x,m_\y,n_\x,n_\y\in\bb Z$. Regularity of the modes $ \phi_{m_\x,m_\y}$ and $B_{n_\x,n_\y}$ at $\eta=0$ requires  $m_\y\geq   r/2$ and $n_\y\leq \p{r-2}/2$.  The   Lagrangian $\L_{\rm chi}'$ that we use as a    $\delta$-exact deformation term  encodes Robin boundary conditions, meaning that  $\hat \L_{\wt P}\,\phi$, $\hat \L_{  P}\,\wt\phi$, $B$ and $\wt B$ have to vanish at  $\eta\to\infty$. The bosonic modes contributing to   $Z_{\rm chi}$ satisfy Robin conditions already  in the bulk of $H^2\times S^1$; thus, they are left unconstrained. On the other hand, the fermionic modes are supposed to vanish  at infinity. This leads us to consider normalizable modes for  $B$, forcing $r<1$. Conversely, Dirichlet conditions at infinity leave $B$ unconstrained and fix $r>1$.  To infer regularity and normalizability  of the fields  we employed  the norm induced by the inner product
 \be{\label{eq: fieldsinnerproduct}
 \langle X_1,X_2\rangle=\int_Md^3x\,\sqrt g\,(X_1)^\ddagger X_2\,.
 }
   Consequently,   the one-loop determinant   (\ref{eq: oneloopdetkernelcokernel}) with Robin boundary conditions is 
 \ba{
    Z_{{\rm chi}}&=\prod_{ n_\x \in\bb Z} \prod_{n_\y\geq 0}\f{  n_\x+\a\, (n_\y-\f{r-2}2)+ \im \,u}{n_\x+\a\, (n_\y+\f r2)- \im \,u} \,,\nn\\
    & =\f{\Gamma_2(\f{\a r}2-\im u|1,\a)\Gamma_2(1-\f{\a r}2+\im u|1,-\a)}{\Gamma_2(\a-\f{\a r}2+\im u|1,\a)\Gamma_2(1-\a+\f{\a r}2-\im u|1,-\a)} =e^{\im\pi\, \mathcal A_{\rm chi}}\f{(t \,q^{1-\f r2};q)}{(t^{-1}\, q^{\f r2};q)}\,,
    }
     with $r>1$, $u = L\,\b\p{\sigma+\im\,K\cdot v}$ as well as $t=e^{2\pi   u}$ and $q=e^{2\pi\im\a}$. The phase factor $\mathcal A_{\rm chi}$ is
     \be{
     \mathcal A_{\rm chi}= \zeta_2(0,\a - \f{\a r}2+\im u|1,\a)-\im\pi\,\zeta_2(0,\f{\a r}2-\im u|1,\a),
     }
     proving (\ref{eq: nonnormresult}). In computing $Z_{{\rm chi}}$ we regularized the infinite products via  Shintani-Barnes  multiple Zeta and Gamma functions. If we   require   all fields to be normalizable according to (\ref{eq: fieldsinnerproduct}), we see that $\phi$ and $B$ cannot contribute to $Z_{\rm chi}$ at the same time.  In particular, $\phi$ modes will generate a non-trivial $Z_\phi$ for $r>1$, whereas  $B$-modes will produce    $Z_B$ for $r<1$. This shows (\ref{eq: secondresult}).

 %%%%%%%%%%%%%%%%%%%%%%
  %%%%%%%%%%%%%%%%%%%%%%

%%%%%%%%%%  Appendices & References    %%%%%%%%%%%

 %%%%%%%%%%%%%%%%%%%%%%
  %%%%%%%%%%%%%%%%%%%%%%

\bibliographystyle{unsrt}

\bibliography{pitt_loc}

\begin{thebibliography}{10}

\bibitem{Witten:1992xu}
Edward Witten.
\newblock {Two-dimensional gauge theories revisited}.
\newblock {\em J. Geom. Phys.}, 9:303--368, 1992.

\bibitem{Nekrasov:2003af}
Nikita~A. Nekrasov.
\newblock {Seiberg-Witten prepotential from instanton counting}.
\newblock In {\em {International Congress of Mathematicians (ICM 2002) Beijing,
  China, August 20-28, 2002}}, 2003.

\bibitem{Nekrasov:2003rj}
Nikita Nekrasov and Andrei Okounkov.
\newblock {Seiberg-Witten theory and random partitions}.
\newblock {\em Prog. Math.}, 244:525--596, 2006.

\bibitem{Pestun:2007rz}
Vasily Pestun.
\newblock {Localization of gauge theory on a four-sphere and supersymmetric
  Wilson loops}.
\newblock {\em Commun. Math. Phys.}, 313:71--129, 2012.

\bibitem{Kapustin:2009kz}
Anton Kapustin, Brian Willett, and Itamar Yaakov.
\newblock {Exact Results for Wilson Loops in Superconformal Chern-Simons
  Theories with Matter}.
\newblock {\em JHEP}, 03:089, 2010.

\bibitem{Marino:2009jd}
Marcos Marino and Pavel Putrov.
\newblock {Exact Results in ABJM Theory from Topological Strings}.
\newblock {\em JHEP}, 06:011, 2010.

\bibitem{Festuccia:2011ws}
Guido Festuccia and Nathan Seiberg.
\newblock {Rigid Supersymmetric Theories in Curved Superspace}.
\newblock {\em JHEP}, 06:114, 2011.

\bibitem{Dumitrescu:2012ha}
Thomas~T. Dumitrescu, Guido Festuccia, and Nathan Seiberg.
\newblock {Exploring Curved Superspace}.
\newblock {\em JHEP}, 08:141, 2012.

\bibitem{Closset:2012ru}
Cyril Closset, Thomas~T. Dumitrescu, Guido Festuccia, and Zohar Komargodski.
\newblock {Supersymmetric Field Theories on Three-Manifolds}.
\newblock {\em JHEP}, 05:017, 2013.

\bibitem{Closset:2013vra}
Cyril Closset, Thomas~T. Dumitrescu, Guido Festuccia, and Zohar Komargodski.
\newblock {The Geometry of Supersymmetric Partition Functions}.
\newblock {\em JHEP}, 01:124, 2014.

\bibitem{Closset:2014uda}
Cyril Closset, Thomas~T. Dumitrescu, Guido Festuccia, and Zohar Komargodski.
\newblock {From Rigid Supersymmetry to Twisted Holomorphic Theories}.
\newblock {\em Phys. Rev.}, D90(8):085006, 2014.

\bibitem{Benini:2015eyy}
Francesco Benini, Kiril Hristov, and Alberto Zaffaroni.
\newblock {Black hole microstates in AdS$_{4}$ from supersymmetric
  localization}.
\newblock {\em JHEP}, 05:054, 2016.

\bibitem{Benini:2016rke}
Francesco Benini, Kiril Hristov, and Alberto Zaffaroni.
\newblock {Exact microstate counting for dyonic black holes in AdS4}.
\newblock {\em Phys. Lett.}, B771:462--466, 2017.

\bibitem{Cabo-Bizet:2017jsl}
Alejandro Cabo-Bizet, Victor~I. Giraldo-Rivera, and Leopoldo~A. Pando~Zayas.
\newblock {Microstate counting of AdS$_{4}$ hyperbolic black hole entropy via
  the topologically twisted index}.
\newblock {\em JHEP}, 08:023, 2017.

\bibitem{Hosseini:2017mds}
Seyed~Morteza Hosseini, Kiril Hristov, and Alberto Zaffaroni.
\newblock {An extremization principle for the entropy of rotating BPS black
  holes in AdS$_{5}$}.
\newblock {\em JHEP}, 07:106, 2017.

\bibitem{Hosseini:2017fjo}
Seyed~Morteza Hosseini, Kiril Hristov, and Achilleas Passias.
\newblock {Holographic microstate counting for AdS$_{4}$ black holes in massive
  IIA supergravity}.
\newblock {\em JHEP}, 10:190, 2017.

\bibitem{Benini:2017oxt}
Francesco Benini, Hrachya Khachatryan, and Paolo Milan.
\newblock {Black hole entropy in massive Type IIA}.
\newblock {\em Class. Quant. Grav.}, 35(3):035004, 2018.

\bibitem{Cabo-Bizet:2018ehj}
Alejandro Cabo-Bizet, Davide Cassani, Dario Martelli, and Sameer Murthy.
\newblock {Microscopic origin of the Bekenstein-Hawking entropy of
  supersymmetric AdS$_{\bf 5}$ black holes}.
\newblock 2018.

\bibitem{Choi:2018hmj}
Sunjin Choi, Joonho Kim, Seok Kim, and June Nahmgoong.
\newblock {Large AdS black holes from QFT}.
\newblock 2018.

\bibitem{Benini:2018ywd}
Francesco Benini and Paolo Milan.
\newblock {Black holes in 4d $\mathcal{N}=4$ Super-Yang-Mills}.
\newblock 2018.

\bibitem{Pestun:2016zxk}
Vasily Pestun et~al.
\newblock {Localization techniques in quantum field theories}.
\newblock {\em J. Phys.}, A50(44):440301, 2017.

\bibitem{Benini:2012ui}
Francesco Benini and Stefano Cremonesi.
\newblock {Partition Functions of ${\mathcal{N}=(2,2)}$ Gauge Theories on
  S$^{2}$ and Vortices}.
\newblock {\em Commun. Math. Phys.}, 334(3):1483--1527, 2015.

\bibitem{Benini:2015noa}
Francesco Benini and Alberto Zaffaroni.
\newblock {A topologically twisted index for three-dimensional supersymmetric
  theories}.
\newblock {\em JHEP}, 07:127, 2015.

\bibitem{Benini:2016hjo}
Francesco Benini and Alberto Zaffaroni.
\newblock {Supersymmetric partition functions on Riemann surfaces}.
\newblock {\em Proc. Symp. Pure Math.}, 96:13--46, 2017.

\bibitem{Alday:2013lba}
Luis~F. Alday, Dario Martelli, Paul Richmond, and James Sparks.
\newblock {Localization on Three-Manifolds}.
\newblock {\em JHEP}, 10:095, 2013.

\bibitem{Closset:2013sxa}
Cyril Closset and Itamar Shamir.
\newblock {The $\mathcal{N}=1$ Chiral Multiplet on $T^2\times S^2$ and
  Supersymmetric Localization}.
\newblock {\em JHEP}, 03:040, 2014.

\bibitem{Assel:2014paa}
Benjamin Assel, Davide Cassani, and Dario Martelli.
\newblock {Localization on Hopf surfaces}.
\newblock {\em JHEP}, 08:123, 2014.

\bibitem{Closset:2015rna}
Cyril Closset, Stefano Cremonesi, and Daniel~S. Park.
\newblock {The equivariant A-twist and gauged linear sigma models on the
  two-sphere}.
\newblock {\em JHEP}, 06:076, 2015.

\bibitem{Festuccia:2016gul}
Guido Festuccia, Jian Qiu, Jacob Winding, and Maxim Zabzine.
\newblock {$ \mathcal{N}=2 $ supersymmetric gauge theory on connected sums of
  $S^{2} \times S^{2}$}.
\newblock {\em JHEP}, 03:026, 2017.

\bibitem{Polydorou:2017jha}
Konstantina Polydorou, Andreas Rocén, and Maxim Zabzine.
\newblock {7D supersymmetric Yang-Mills on curved manifolds}.
\newblock {\em JHEP}, 12:152, 2017.

\bibitem{Festuccia:2018rew}
Guido Festuccia, Jian Qiu, Jacob Winding, and Maxim Zabzine.
\newblock {Twisting with a Flip (the Art of Pestunization)}.
\newblock 2018.

\bibitem{Hori:2013ika}
Kentaro Hori and Mauricio Romo.
\newblock {Exact Results In Two-Dimensional (2,2) Supersymmetric Gauge Theories
  With Boundary}.
\newblock 2013.

\bibitem{Yoshida:2014ssa}
Yutaka Yoshida and Katsuyuki Sugiyama.
\newblock {Localization of 3d $\mathcal{N}=2$ Supersymmetric Theories on $S^1
  \times D^2$}.
\newblock 2014.

\bibitem{Gava:2016oep}
Edi Gava, K.~S. Narain, M.~Nouman Muteeb, and V.~I. Giraldo-Rivera.
\newblock {$N = 2$ gauge theories on the hemisphere $HS^4$}.
\newblock {\em Nucl. Phys.}, B920:256--297, 2017.

\bibitem{Bawane:2017gjf}
Aditya Bawane, Sergio Benvenuti, Giulio Bonelli, Nouman Muteeb, and Alessandro
  Tanzini.
\newblock {$\mathcal{N}=2$ gauge theories on unoriented/open four-manifolds and
  their AGT counterparts}.
\newblock 2017.

\bibitem{Aharony:2015hix}
Ofer Aharony, Micha Berkooz, Avner Karasik, and Talya Vaknin.
\newblock {Supersymmetric field theories on AdS$_{p} \times$ S$^{q}$}.
\newblock {\em JHEP}, 04:066, 2016.

\bibitem{Bonetti:2016nma}
Federico Bonetti and Leonardo Rastelli.
\newblock {Supersymmetric localization in AdS$_{5}$ and the protected chiral
  algebra}.
\newblock {\em JHEP}, 08:098, 2018.

\bibitem{David:2016onq}
Justin~R. David, Edi Gava, Rajesh~Kumar Gupta, and Kumar Narain.
\newblock {Localization on AdS$_{2} \times$ S$^{1}$}.
\newblock {\em JHEP}, 03:050, 2017.

\bibitem{Assel:2016pgi}
Benjamin Assel, Dario Martelli, Sameer Murthy, and Daisuke Yokoyama.
\newblock {Localization of supersymmetric field theories on non-compact
  hyperbolic three-manifolds}.
\newblock {\em JHEP}, 03:095, 2017.

\bibitem{David:2018pex}
Justin~R. David, Edi Gava, Rajesh~Kumar Gupta, and Kumar Narain.
\newblock {Boundary conditions and localization on AdS. Part I}.
\newblock {\em JHEP}, 09:063, 2018.

\bibitem{Feng:2007ur}
Bo~Feng, Amihay Hanany, and Yang-Hui He.
\newblock {Counting gauge invariants: The Plethystic program}.
\newblock {\em JHEP}, 03:090, 2007.

\bibitem{Almuhairi:2011ws}
Ahmed Almuhairi and Joseph Polchinski.
\newblock Magnetic ads$ \times { R}^2$: Supersymmetry and stability.
\newblock 2011.

\bibitem{Kutasov:2013ffl}
David Kutasov and Jennifer Lin.
\newblock {(0,2) Dynamics From Four Dimensions}.
\newblock {\em Phys. Rev.}, D89(8):085025, 2014.

\bibitem{Beem:2012mb}
Christopher Beem, Tudor Dimofte, and Sara Pasquetti.
\newblock {Holomorphic Blocks in Three Dimensions}.
\newblock {\em JHEP}, 12:177, 2014.

\bibitem{Nieri:2015yia}
Fabrizio Nieri and Sara Pasquetti.
\newblock {Factorisation and holomorphic blocks in 4d}.
\newblock {\em JHEP}, 11:155, 2015.

\bibitem{Imamura:2011su}
Yosuke Imamura and Shuichi Yokoyama.
\newblock {Index for three dimensional superconformal field theories with
  general R-charge assignments}.
\newblock {\em JHEP}, 04:007, 2011.

\bibitem{Dimofte:2017tpi}
Tudor Dimofte, Davide Gaiotto, and Natalie~M. Paquette.
\newblock {Dual boundary conditions in 3d SCFT’s}.
\newblock {\em JHEP}, 05:060, 2018.

\bibitem{Hama:2011ea}
Naofumi Hama, Kazuo Hosomichi, and Sungjay Lee.
\newblock {SUSY Gauge Theories on Squashed Three-Spheres}.
\newblock {\em JHEP}, 05:014, 2011.

\end{thebibliography}

\end{document}